\documentstyle[epsfig]{l-aa}  
\def\grs{\mbox{GRS } 1915+105}
\def\gcloud{\mbox{G } 45.46+0.06}
\def\ss{\mbox{SS } 433}
\def\cygtrois{\mbox{Cyg X-} 3}
\def\cygun{\mbox{Cyg X-} 1}
\def\hun{\mbox{H\,{\sc i}}}
\def\hdeux{\mbox{H\,{\sc ii}}}
\def\lambdavingtetun{\lambda \, 21 \mbox{ cm}}
\def\co{^{12} \mbox{CO } (\mbox{J} = 1 - 0)}
\def\nh{N \mbox{(H)}}
\def\plusoumoins{\, \pm \,}
\def\kpc{\mbox{ kpc}}
\def\ergs{\mbox{ erg s}^{-1}}
\def\cmdeux{\mbox{ cm}^{-2}}
\def\cmtrois{\mbox{ cm}^{-3}}
\def\Av{A_{\rm v}}
\def\kms{\mbox{ km s}^{-1}}
\def\Msun{\mbox{ }M_{\odot}}

\def\Kkmsdeg2{\mbox{ K km s}^{-1} \mbox{deg}^{2}}
\def\micrometre{\mbox{ } \mu \mbox{m}}
\def\mag{\mbox{ magnitude}}
\def\mags{\mbox{ magnitudes}}
\def\K{\mbox{ K}}

\begin{document}
\thesaurus{
08      
        (08.09.2 GRS 1915+105; 
         08.22.3)              
09      
        (09.03.1)              
13      
        (13.07.2;              
         13.09.6;              
         13.25.5)              
}

\title{Infrared and millimeter observations of the galactic superluminal source $\grs$}

\author{S. Chaty \inst{1} \and I.F. Mirabel \inst{1} \and P.A. Duc \inst{1} \and J.E. Wink \inst{2} \and L.F. Rodr\'{\i}guez \inst{3}}

\offprints{S. Chaty (chaty@discovery.saclay.cea.fr) or I.F. Mirabel (mirabel@discovery.saclay.cea.fr)}

\institute{
Service d'Astrophysique, CEA/DSM/DAPNIA/SAp, Centre d'\'etudes de Saclay, F-91191 Gif-sur-Yvette Cedex, France
\and
IRAM, 300, rue de la Piscine, Domaine Universitaire de Grenoble, F-38406 Saint-Martin-d'H\`eres Cedex, France
\and
Instituto de Astronom\'{\i}a, UNAM, Apdo Postal 70-264, M\'exico, DF, 04510, Mexico
}

\date{Received 16 August 1995; accepted 13 November 1995}

\maketitle

\markboth{S. Chaty et al.: Infrared and millimeter observations of $\grs$}{}

\begin{abstract}
Millimeter observations of the galactic source of relativistic ejections $\grs$ (Mirabel \& Rodr\'{\i}guez 1994) are consistent with this source being at a kinematic distance $D = 12.5 \plusoumoins 1.5 \kpc$ from the Sun, behind the core of a molecular cloud at $9.4 \plusoumoins 0.2 \kpc$. At this distance, $\grs$, frequently radiating $\sim 3 \times 10^{38} \ergs$ in the X-rays, becomes the most luminous X-ray source in the Galaxy. The total hydrogen column density $\nh = 4.7 \plusoumoins 0.2 \times 10^{22} \cmdeux$ along the line of sight corresponds to a visual absorption $\Av = 26.5 \plusoumoins 1 \mags$.

The infrared counterpart of $\grs$ exhibits in the $1.2 \micrometre$ -- $2.2 \micrometre$ band variations of $\sim 1 \mag$ in a few hours and of $\sim 2 \mags$ over longer intervals of time. In the infrared, $\grs$ is strikingly similar to $\ss$, and unlike any other known stellar source in the Galaxy. The infrared resemblance in absolute magnitude, color, and time variability, between these two sources of relativistic ejections suggests that $\grs$, as $\ss$, consists of a collapsed object (neutron star or black hole) with a thick accretion disk in a high-mass-luminous binary system.


\end{abstract}

\section{Introduction}

Since its discovery on 15 August 1992 by the WATCH all-sky X-ray monitor on board of GRANAT (Castro-Tirado et al. 1992), the hard X-ray transient $\grs$ in the constellation of Aquila, has shown long periods of violent erratic variations with recurrent rises to maximum luminosity (Harmon et al. 1992; Brandt et al. 1993; Castro-Tirado et al. 1994). The fairly hard spectrum with emission up to 220 keV and variable spectral index between $-2$ and $-2.8$ observed by BATSE on the Compton Gamma-Ray Observatory (GRO) indicate that $\grs$ is a collapsed object, likely a black hole in a binary system (Harmon et al. 1994). The arcmin location by SIGMA on GRANAT (Finoguenov et al. 1994) allowed the detection of variable radio and infrared counterparts (Mirabel et al. 1994). Follow-up observations with the VLA at centimeter wavelengths led to the discovery of relativistic ejections of plasma clouds with apparent superluminal motions (Mirabel \& Rodr\'{\i}guez 1994, 1995; Rodr\'{\i}guez \& Mirabel 1995).\\

Here we report the results from a multi-wavelength approach to determine the distance and environment of this unique X-ray source. The millimeter and centimeter wavelengths observations yield constraints on its kinematic distance and give clues on the environment of the source. This source is located near the galactic plane at $\mbox{\it l} = 45.4 \degr$, $\mbox{\it b} = -0.3 \degr$ and due to interstellar extinction it is better observed in the infrared (Mirabel et al. 1994). To constrain the nature of the binary system, imaging and spectroscopic observations were carried out in the J ($1.25 \micrometre$), H ($1.65 \micrometre$), and K ($2.2 \micrometre$) bands. The comparison between this source and other well known X-ray binaries is also discussed.

\section{Interstellar gas in the direction to $\grs$}

\subsection{Observations of the atomic hydrogen}

\begin{figure}
\centerline{\psfig{file=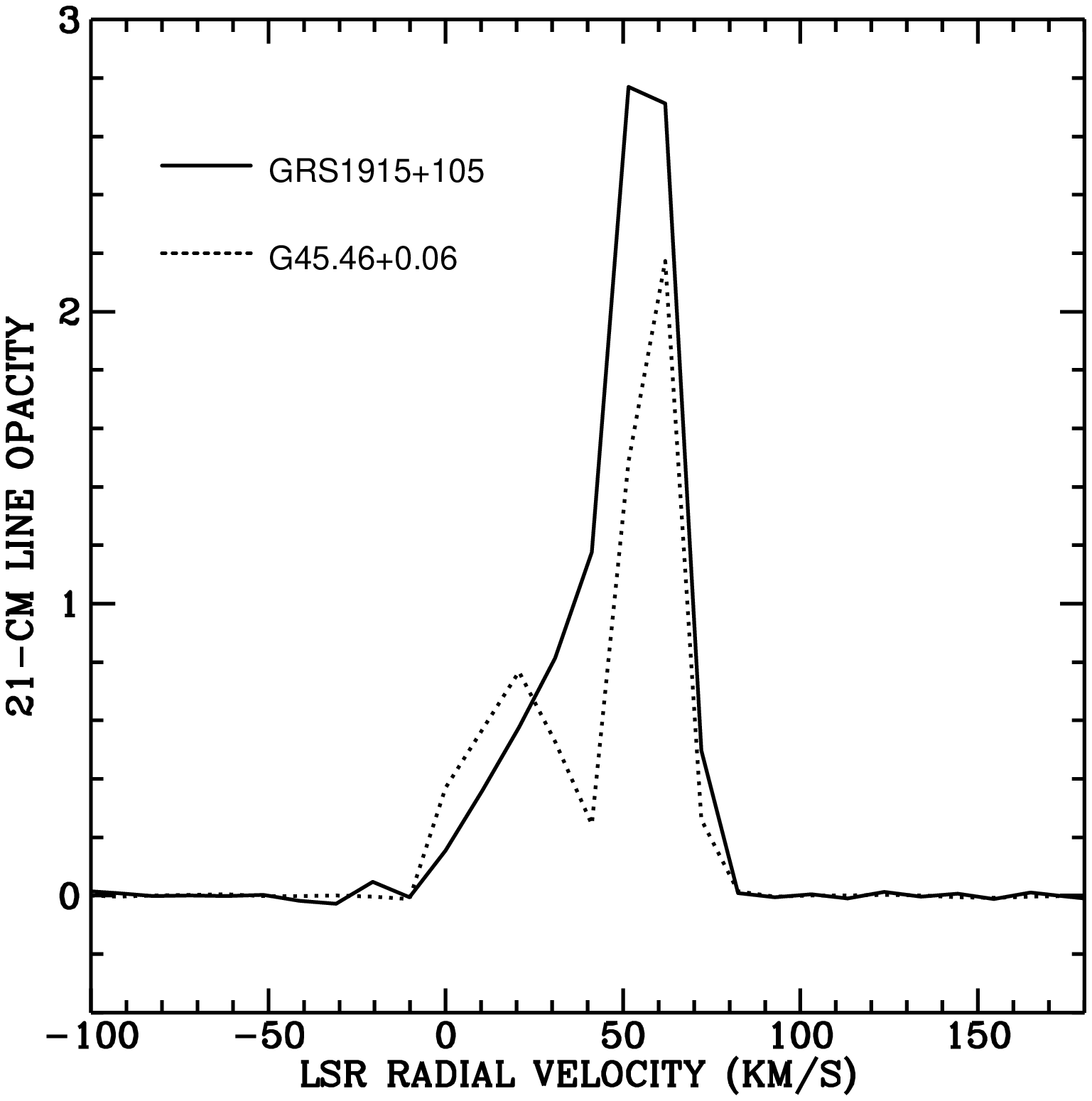,angle=0.,width=8.7cm}} 
\caption[]{\label{GRSabs} Opacities of $\lambdavingtetun$ line from atomic hydrogen absorption spectra observed along the lines of sight to $\grs$ and to the $\hdeux$ region $\gcloud$ (Rodr\'{\i}guez et al. 1995) with the VLA in configuration D on 16 December 1993. The resolution is 10.3 km.s$^{-1}$. Note that the line of sight to $\grs$ exhibits a relative excess of $\hun$ absorption at LSR radial velocity $41 \plusoumoins 6 \kms$.}
\end{figure}

Using the Very Large Array (VLA) in configuration D on 16 December 1993, Rodr\'{\i}guez et al. (1995) observed with a resolution of $10.3 \kms$ the $\lambdavingtetun$ line absorption spectra of $\hun$ along the lines of sight to $\grs$ and to the $\hdeux$ region $\gcloud$. The latter is the closest $\hdeux$ region on the sky at a projected distance of $17 \mbox{ arcmin}$ from the high-energy source. Its radial velocity with respect to the local standard of rest (LSR) is $+53.6 \plusoumoins 1 \kms$. The kinematic distance of this $\hdeux$ region can be unambiguously known combining the $\lambdavingtetun$ absorption with the H109$\alpha$ and H110$\alpha$ emission (Downes et al. 1980; Matthews et al. 1977; Baud 1977). Assuming that the Sun is at a Galactocentric distance of $8.5 \kpc$, the $\hdeux$ region is at $\sim 8.8 \kpc$ from the Sun. The $\lambdavingtetun$ opacities toward $\grs$ are shown in Fig. \ref{GRSabs}. Since there is absorption in both the $\hdeux$ region and $\grs$ up to  $70 \kms$ (which corresponds to the LSR velocity of the subcentral point for the galactic longitude $45 \fdg 4$), both the $\hdeux$ region and $\grs$ are beyond $6 \kpc$ from the Sun (Radhakrishnan et al. 1972).\\

The line of sight to $\grs$ exhibits additional $\hun$ absorption at $41 \plusoumoins 6 \kms$ relative to the line of sight to $\gcloud$, implying that this additional $\hun$ absorption is farther than the $\hdeux$ region $\gcloud$. So $\grs$ appears to be behind an $\hun$ cloud, which is at a kinematic distance of $9.4 \plusoumoins 0.5 \kpc$, beyond $\gcloud$. On the other hand, since there is no absorption at negative velocities below $-25 \kms$, $\grs$ must be at a kinematic distance $\leq 14 \kpc$, which is consistent with the relativistic upper limit of the distance derived by Mirabel \& Rodr\'{\i}guez (1994). Therefore we can constrain the kinematic distance of $\grs$ to the range $7.9 \leq \mbox{D (kpc)} \leq 14$. The $\hun$ absorption denotes a column density of atomic gas $N(\hun) = (1.73 \times 10^{22})(T_{\rm s}/100\mbox{ K}) \cmdeux$, where $T_{\rm s}$ is the spin temperature. This column density is 1.42 times the column density along the line of sight to the $\hdeux$ region $\gcloud$. Assuming a constant $\hun$ absorption per unit length, $\grs$ would then be at a kinematic distance $D = 12.5 \plusoumoins 1.5 \kpc$ (Rodr\'{\i}guez et al. 1995).

\subsection{Observations of the molecular gas}

Low resolution observations of the interstellar molecular gas have been discussed by Castro-Tirado et al. (1994) and Grindlay (1994). Using the Columbia CO survey (Dame et al. 1986), they found that along the line of sight to $\grs$ there are molecular gas complexes at the kinematic distances of 1.5 and 7.8 kpc. However, due to the low angular resolution of the Columbia survey, no molecular cloud core was detected on the line of sight, and therefore no evidence for association between the source and a giant molecular cloud was found.\\

We used the 30~m radiotelescope of the Institut de Radioastronomie Millim\'etrique (IRAM) for a search in the direction of $\grs$ of molecular gas associated with the additional $\hun$ absorption detected with the VLA at $41 \plusoumoins 6 \kms$. The observations were made from 12 to 19 October 1994, with the 3 mm-receiver, at the rest frequency 115.271204 GHz of the $\co$ transition. This transition is a good density tracer, for molecular hydrogen densities of $100 - 300 \cmtrois$ (Sanders et al. 1983). We chopped against an off region, which was conveniently chosen, so that it contains no feature. The coordinates of this reference off region are: $ \alpha(1950) = 19^{\rm h}06^{\rm m}38 \fs 2 $, $ \delta(1950) = 11 \degr 03 \arcmin 39 \arcsec $.\\

\begin{figure}
\centerline{\psfig{file=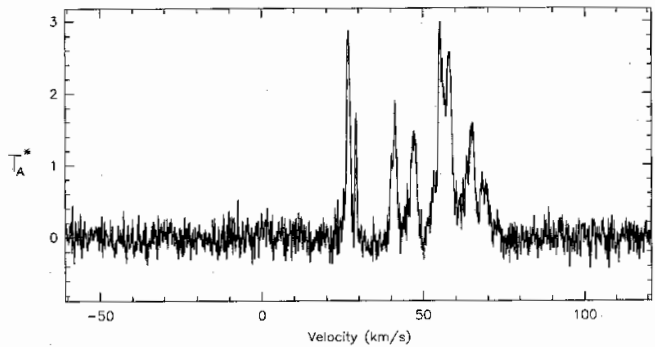,angle=89.5,width=8.8cm}} 
\caption[]{\label{COprofile} $\co$ emission in the direction of $\grs$ observed with the IRAM 30~m telescope.}
\end{figure}

Figure \ref{COprofile} shows the $\co$ spectrum toward $\grs$. There is an emission peak of the CO at the velocity $41.5 \plusoumoins 1 \kms$, close to the LSR velocity of the atomic component that causes the additional absorption in $\hun$. In the following we assume that the $\hun$ absorption detected with the VLA at the velocity $41 \plusoumoins 6 \kms$ and the CO feature detected with the 30~m telescope of the IRAM at $41.5 \plusoumoins 1 \kms$ come from the same cloud. Therefore we adopt the far distance for the CO component, and its Doppler shift corresponds to a kinematic distance of $9.4 \plusoumoins 0.2 \kpc$. The velocity width of the molecular cloud is $1.9 \plusoumoins 0.1 \kms$.\\

\begin{figure}
\centerline{\psfig{file=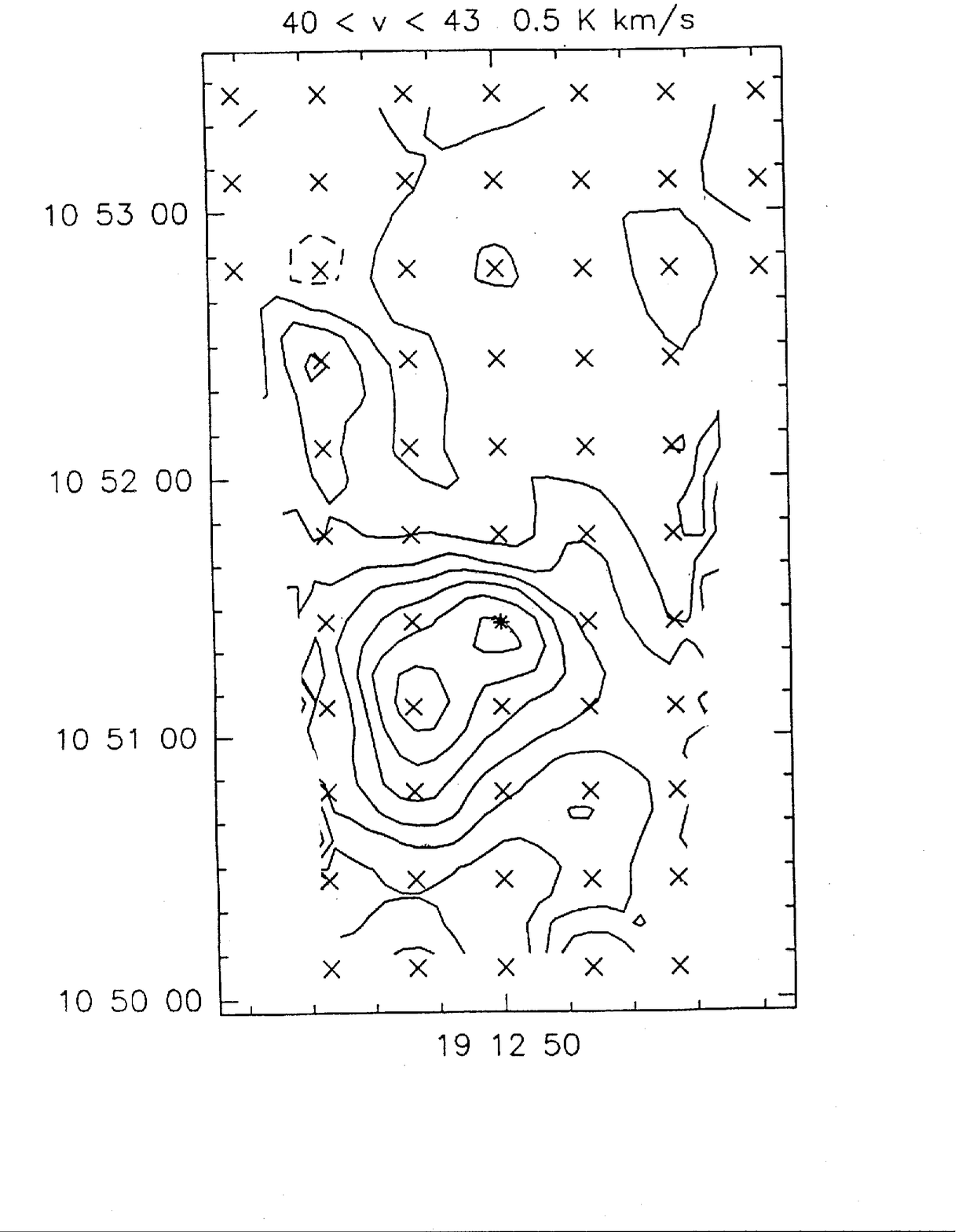,angle=-1,width=10cm}} 
\caption[]{\label{COmap} Map of the $\co$ emission for the LSR velocity $40 \leq \, v \, \leq 43 \kms$ in the direction of $\grs$. The asterisk indicates the position of the VLA radio counterpart of the high-energy source. The contours are -0.5, 0.5 to 3.5 by 0.5, in units of $K \, T_{\rm A}^{\star} \kms$ for a LSR velocity interval of 40 to 43 km.s$^{-1}$. The observed positions are represented by crosses. Coordinates are in the B1950 equinox.}
\end{figure}

To study the distribution of the molecular gas at given Doppler shifts, we made maps of the CO emission with velocities covering the following intervals in km.s$^{-1}$: ($-10$,\,0), (0,\,10), (23,\,30), (26,\,28), (28,\,30), (40,\,43), (43,\,46), (46,\,50), (57,\,60), (60,\,62), and (62,\,70). With these intervals of velocities we mapped the CO emission with kinematic distances as far as $12.8 \kpc$. The only map that shows a local maximum of the CO emission near the projected position of $\grs$ is the map of the CO detected in the range of velocities between 40 and $43 \kms$, centered at $41.5 \kms$. This map is shown in Fig. \ref{COmap}. From it we can conclude that $\grs$ is behind the core of a molecular cloud located at the kinematic distance $D = 9.4 \plusoumoins 0.2 \kpc$ from the Sun.\\

The molecular cloud superimposed upon the position of $\grs$ has a diameter of $1.05 \mbox{ arcmin}$, which corresponds to a physical size of 2.9 pc for a kinematic distance of $9.4 \kpc$. Thus, the mean density of the molecular cloud is $\sim 400 \cmtrois$. One can derive the mass of the cloud using the equation $M(\mbox{H}_2) = 1.3 \times 10^{3} \, S \, D^{2}$, where $S = 1.24 \times 10^{-2} \Kkmsdeg2$ is the apparent CO luminosity of the cloud (Dame et al. 1986). The cloud mass is $M(\mbox{H}_2) = 1400 \plusoumoins 100 \Msun$. The column density of molecular hydrogen H$_{2}$ is $N(\mbox{H}_{2}) = 3.6 \times 10^{20} \int \, T_{\rm A}^{\star} \mbox{(CO) }dv = 1.5 \plusoumoins 0.1 \times 10^{22} \cmdeux$, where the integrated antenna temperature is $T_{\rm A}^{\star}(\mbox{CO}) = 31.027 \mbox{ K}$, and the averaged velocity width is $dv = 1.3 \kms$. So the total hydrogen column density is $\nh = N(\hun)+2N(\mbox{H}_{2}) = 4.7 \plusoumoins 0.2 \times 10^{22} \cmdeux$. This is consistent, within the errors, with the column density of $\sim 5 \times 10^{22} \cmdeux$ derived from the ROSAT X-ray spectrum of the source (Greiner et al. 1994). The visual absorption derived from the column density, using the equation $\Av \mbox{(mag)} = 5.59 \times 10^{-22} \nh (\cmdeux)$ (Predehl \& Schmitt 1995), is $\Av = 26.5 \plusoumoins 1 \mags$.

\section{The infrared counterpart of $\grs$}

\subsection{Infrared observations}

Mirabel et al. (1994) had shown that there is no visual counterpart of $\grs$ brighter than $R = 21 \mags$. Using the NTT with a Gunn-z filter on 9 July 1994, we observed at $\sim 1 \micrometre$ a faint counterpart consistent with the $I = 23.4 \mags$ counterpart reported by Bo\"er et al. (1995).\\

We carried out infrared observations of $\grs$ at the European Southern Observatory (ESO) with the ESO/MPI 2.2~m telescope on 4 and 5 June 1993 with the IRAC2 camera (Mirabel et al. 1994), and from 5 to 8 July 1994 with the IRAC2b camera in the J ($1.25 \micrometre$), H ($1.65 \micrometre$) and K ($2.2 \micrometre$) bands. The IRAC2(b) camera was mounted at the F/35 infrared adapter of the telescope. This camera is a Rockwell 256$\times$256 pixels Hg:Cd:Te NICMOS 3 large format infrared array detector. It was used with the lens C, providing an image scale of $0.49 \mbox{ arcsec/pixel}$ and a field of $136 \times 136 \mbox{ arcsec}^{2}$. The typical seeing for these observations was $1.2 \mbox{ arcsec}$.\\

Follow-up observations were performed on our request by S. Massey at the 3.6~m Canada-France-Hawaii Telescope (hereafter CFHT) on Mauna Kea on 16 August 1994, with the Redeye camera, in the H and K bands. The narrow field infrared Redeye camera was mounted at the F/8 focus of the CFHT. This camera is a Rockwell 256$\times$256 pixels Hg:Cd:Te NICMOS 3 infrared array detector, providing a plate scale of $0.20 \mbox{ arcsec/pixel}$. The typical seeing for these observations was $0.6 \mbox{ arcsec}$.\\

Each image taken at la Silla is the median of 5 images exposed during 2 minutes. After taking each image of the object, an image of the sky was taken, to allow subtraction of the blank sky. For the images acquired by the CFHT, the result is the median of 18 images exposed during 30 s. The images are further treated by removal of the bias, the dark current, and the flat field, and we carried out absolute and relative photometry, to look for small variations of the luminosity of $\grs$. This work was performed with the IRAF procedures, using the DAOPHOT package for the photometry in crowded fields.\\

\begin{figure}
\centerline{\psfig{file=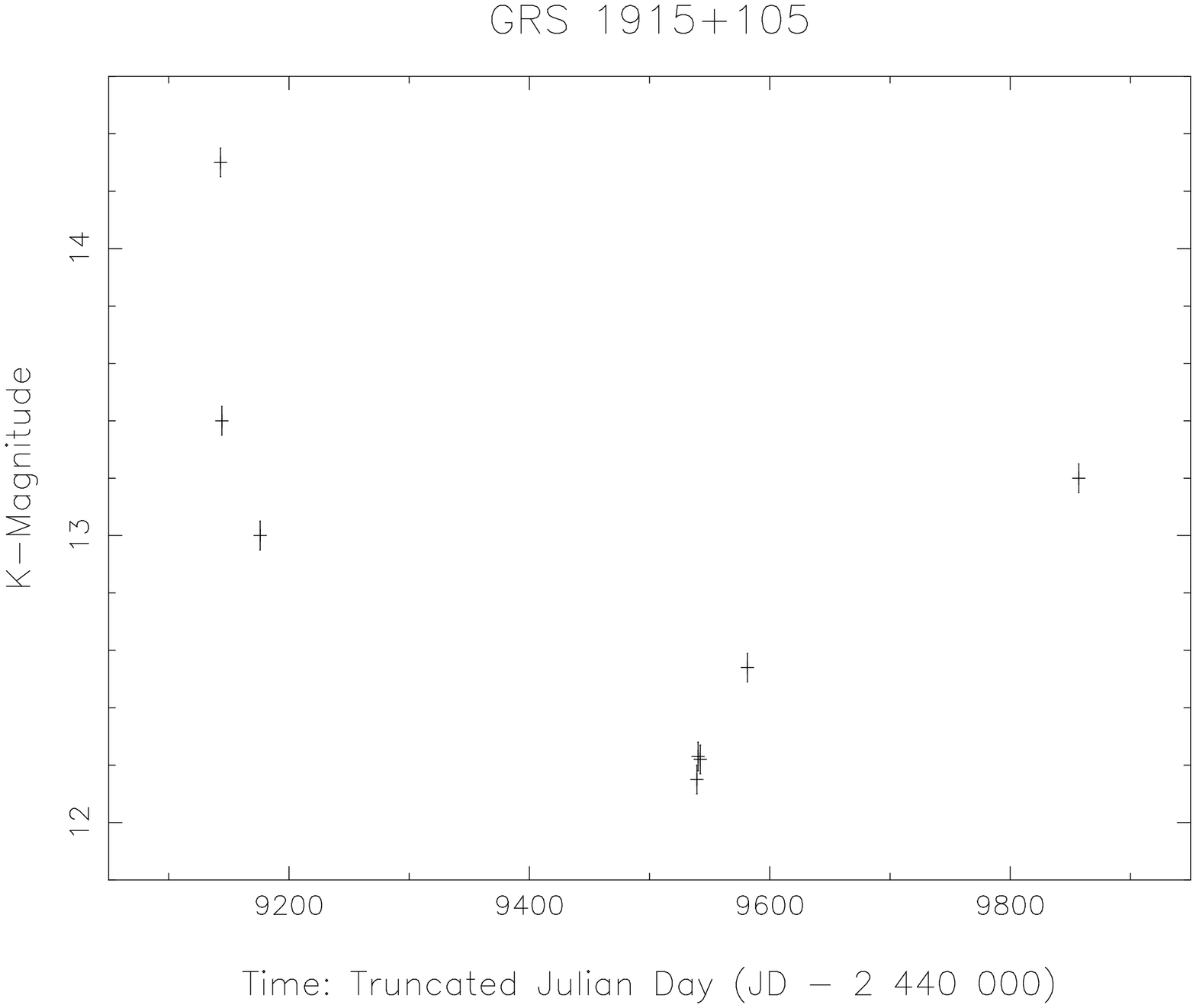,angle=-90,width=10.cm}} 
\caption[]{\label{GRSvarie} Time variation of K-band ($2.2 \micrometre$) luminosity of the source $\grs$ from 4 June 1993 until 19 May 1995.}
\end{figure}

The magnitudes are given in Table \ref{varmag}, and the variations in the K-band are shown in Fig. \ref{GRSvarie}. We can see that $\grs$ exhibits strong variability in the J, H and K bands. The luminosity of $\grs$ increased by nearly 1 magnitude in H and K between the nights of 4 and 5 June 1993. Between 4 June 1993 and 5 July 1994 there was a change by nearly $2 \mags$ in J, $2.5 \mags$ in H, and $2.1 \mags$ in K. However, no variations greater that $0.1 \mags$ were detected in the period 5--8 July 1994 (we took 1 image each 30 minutes during 5 hours, on 5 and 6 July 1994). From Table \ref{varmag} it seems that the infrared colors change with luminosity.\\

\begin{table*}
\caption[]{\label{varmag} {\bf Optical and infrared magnitudes of GRS 1915+105.}}
\begin{flushleft}
\begin{tabular}{llllllll} 
\hline \noalign{\smallskip}
{\bf Date} & {\bf TJD$^{1}$} & {\bf telescope} & {\bf R}($0.7 \micrometre$) & {\bf I}($0.9 \micrometre$) & {\bf J}($1.25 \micrometre$) & {\bf H}($1.65 \micrometre$) & {\bf K}($2.2 \micrometre$)\\
\noalign{\smallskip}
\hline \noalign{\smallskip}
19/04/93 & 9097 & Pic Midi & $>21$ & - & - & - & - \\
04/06/93 & 9143 & ESO 2.2m & - & - & $\geq 18 \plusoumoins 0.2$ & $16.2 \plusoumoins 0.2$ & $14.3 \plusoumoins 0.2$ \\
05/06/93 & 9144 & ESO 2.2m & - & - & $18 \plusoumoins 0.1$ & $15.0 \plusoumoins 0.1$ & $13.4 \plusoumoins 0.1$ \\
27/06/93$^{2}$ & 9166 & CFHT & $>26.1$ & $23.4 \plusoumoins 0.3$ & - & - & - \\
07/07/93$^{3}$ & 9176 & UKIRT & - & - & $16.6 \plusoumoins 0.1$ & - & $13.0 \plusoumoins 0.1$ \\
05/07/94 & 9539 & ESO 2.2m & - & - & $16.2 \plusoumoins 0.1$ & $13.7 \plusoumoins 0.1$ & $12.15 \plusoumoins 0.08$ \\
06/07/94 & 9540 & ESO 2.2m & - & - & - & - & $12.23 \plusoumoins 0.04$ \\
07/07/94 & 9541 & ESO 2.2m & - & - &  - & - & $12.23 \plusoumoins 0.1$ \\
08/07/94 & 9542 & ESO 2.2m & - & - & - & - & $12.22 \plusoumoins 0.1$ \\
09/07/94 & 9543 & NTT & $>22$ & - & - & - & - \\
16/08/94 & 9581 & CFHT & - & - & - & $14.83 \plusoumoins 0.1$ & $12.54 \plusoumoins 0.1$ \\
19/05/95$^{4}$ & 9857 & UKIRT & - & - & - & - & $13.2 \plusoumoins 0.1$ \\
\noalign{\smallskip}
\hline
\end{tabular}
\end{flushleft}
{\footnotesize $^{1}$Truncated Julian Date (JD - 2 440 000).}\\
{\footnotesize $^{2}$From Bo\"er et al. (1995).}\\
{\footnotesize $^{3}$From Castro-Tirado et al. (1993).}\\
{\footnotesize $^{4}$From Geballe (1995).}
\end{table*}

Figure \ref{GRSvarie} shows that $\grs$ exhibits short-term variability in intervals of less than 24 hours as well as long-term variability over intervals from one month to one year. The rapid increase of 1 magnitude observed in an interval of 24 hours in June 1993 could result from occultation. It is also interesting to note that this rapid variation of the infrared luminosity occurred in a period when the source was strong and showing rapid variations of luminosity in the 8--60 keV energy band observed by WATCH (Sazonov et al. 1994), and in the 20--100 keV energy band observed by BATSE (Harmon et al. 1994; Paciesas et al. 1995). The luminous infrared phase of $\grs$ observed in July 1994 corresponds to a period of erratic variations in the X-rays, when the source was fainter in the 20--100 keV energy band than in the period June 1993 (Harmon et al. 1994; Paciesas et al. 1995).\\

\subsection{Discussion}

From the apparent magnitudes in Table \ref{varmag} we derived the absolute magnitudes, corrected for interstellar extinction, using a visual absorption of $\Av = 26.5 \plusoumoins 1\mags$, and the kinematic distance $D = 12.5 \plusoumoins 1.5 \kpc$. The absorptions in the J, H, and K bands are respectively $A_{\rm J} = 7.1 \plusoumoins 0.2$, $A_{\rm H} = 4.1 \plusoumoins 0.2$, and $A_{\rm K} = 3.0 \plusoumoins 0.1 \mags$. The infrared absolute magnitudes and colors of $\grs$ and other well known X-ray galactic sources during minimum and maximum luminosity are listed in Table \ref{grsss}. To derive the absolute magnitudes of $\ss$ we assumed the kinematic distance of $4.2 \plusoumoins 0.5 \kpc$ (van Gorkom et al. 1982) and a visual absorption $\Av = 7.25 \plusoumoins 0.25 \mags$ (McAlary \& McLaren 1980). In view that the distance to $\ss$ is somewhat uncertain, in the following we use its kinematic distance derived from $\hun$ absorption, namely, using the same method used to derive the distance to $\grs$ (Mirabel \& Rodr\'{\i}guez 1994). The estimated errors of the absolute magnitudes in Table \ref{grsss} take into account the uncertainties on the distance and interstellar absorption.\\

\begin{table*}
\caption[]{\label{grsss} {\bf J, H, and K absolute magnitudes of X-ray sources.}}
\begin{flushleft}
\begin{tabular}{llllllll} 
\hline \noalign{\smallskip}
{\bf Source} & {\bf J}($1.25 \micrometre$) & {\bf H}($1.65 \micrometre$) & {\bf K}($2.2 \micrometre$) & {\bf J-K} & {\bf H-K} & {\bf J-H} & ref \\
\noalign{\smallskip} \hline \noalign{\smallskip}
$\grs$ & $-4.5 \plusoumoins 0.9$ & $-4.6 \plusoumoins 0.9$ & $-5.1 \plusoumoins 0.8$ & $0.6$ & $0.5$ & $0.1$ & 1 \\
 & $-6.3 \plusoumoins 0.9$ & $-5.8 \plusoumoins 0.9$ & $-6.4 \plusoumoins 0.8$ & $0.1$ & $0.6$ & $-0.5$ & 2 \\
\hline
$\ss$ & $-5.2 \plusoumoins 0.8$ & $-5.0 \plusoumoins 0.7$ & $-5.5 \plusoumoins 0.7$ & $0.3$ & $0.5$ & $-0.2$ & 3 \\
 & $-6.4 \plusoumoins 0.8$ & $-6.4 \plusoumoins 0.7$ & $-6.8 \plusoumoins 0.7$ & $0.4$ & $0.4$ & $0.0$ & \\
\hline
$\cygtrois$ & $-3.3 \plusoumoins 0.3$ & $-4.1 \plusoumoins 0.3$ & $-4.9 \plusoumoins 0.3$ & $1.6$ & $0.8$ & $0.8$ & 4,5,6 \\
 & $-3.7 \plusoumoins 0.3$ & $-4.8 \plusoumoins 0.3$ & $-5.7 \plusoumoins 0.3$ & $2$ & $0.9$ & $1.1$ & \\
\hline
$\cygun$ & $-5.9 \plusoumoins 0.2$ & $-5.8 \plusoumoins 0.2$ & $-5.8 \plusoumoins 0.2$ & $-0.1$ & $0.0$ & $-0.1$ & 7,8 \\
 & $-6.4 \plusoumoins 0.2$ & - & $-6.3 \plusoumoins 0.2$ & $-0.1$ & - & - & \\
\noalign{\smallskip}
\hline
\end{tabular}
\end{flushleft}
\footnote\footnotesize{see Table \ref{varmag} on 05 June 1993} \\
\footnote\footnotesize{see Table \ref{varmag} on 05 July 1994} \\
\footnote\footnotesize{Catchpole et al. 1981} \\
\footnote\footnotesize{Joyce 1990} \\
\footnote\footnotesize{Jones et al. 1994} \\
\footnote\footnotesize{Becklin et al. 1972} \\
\footnote\footnotesize{Leahy \& Ananth 1992} \\
\footnote\footnotesize{Beall et al. 1984}
\end{table*}

The infrared emission of $\grs$ cannot arise {\bf only} in the photosphere of the secondary star: 1) because of the shape of the spectrum, which cannot be reproduced by photospheric emission from any stellar type (e.g. Koornneef 1983), and 2) because of the rapid variations in luminosity and energy distribution (see Table \ref{varmag}). Therefore, besides the photospheric emission from the secondary, there must be in $\grs$ an additional source of infrared emission.\\

\begin{figure*}
\centerline{\psfig{file=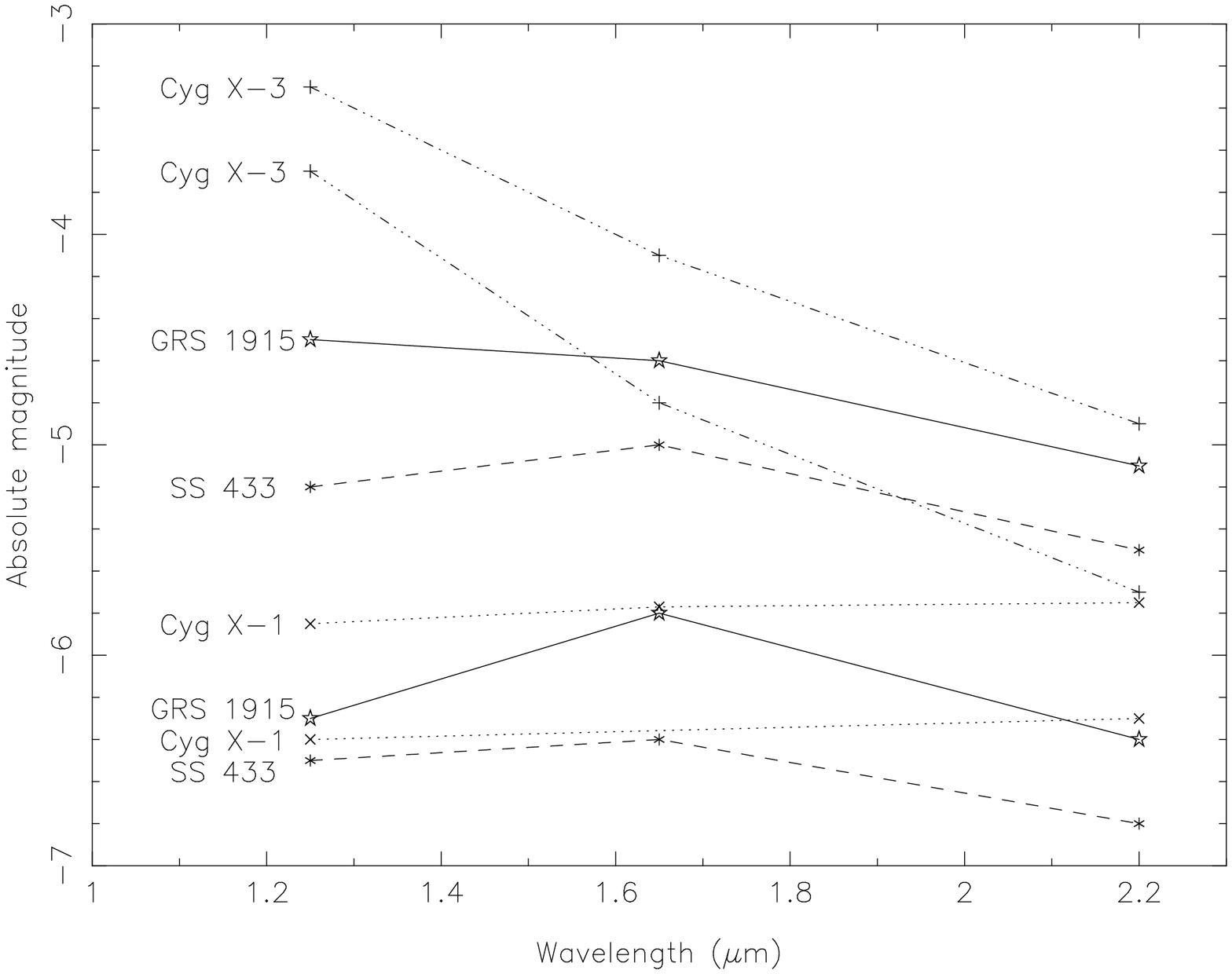,angle=-90.,width=15.cm}} 
\caption[]{\label{grsm143} Infrared energy distributions of $\grs$, $\ss$, $\cygun$ and $\cygtrois$ for the periods of minimum and maximum luminosity.}
\end{figure*}

The energy distribution of the most well studied X-ray galactic sources is shown in Fig. \ref{grsm143}. Besides the time variations, the infrared absolute magnitudes and colors of $\grs$ are strikingly similar to the classic source of relativistic jets $\ss$ (Margon 1984). This similarity in the observed infrared properties suggests that $\ss$ and $\grs$ are systems of similar nature. The infrared emission of $\ss$ arises in a high-mass binary of type late O or early B (Clark \& Milone 1980; Wynn-Williams \& Becklin 1979) or Be (Campbell \& Thompson 1977), with possible contributions of free-free emission from an ionized plasma at $T \sim 7\,500 \K$ (McAlary \& McLaren 1980), an accretion disk (Kodaira et al. 1985), and/or even the jets (Catchpole et al. 1981).\\

Within the context of a binary model with an accretion disk, Kodaira et al. (1985) conclude that the observed infrared flux in the $\ss$ system comes mostly from an accretion disk around the compact object of the binary system, and that the day-to-day variations may be due to different configurations of disk structures, depending on the mass supply and the internal magnetohydrodynamic balances. Therefore, by analogy with $\ss$ (Margon 1984), $\grs$ would be a collapsed object with a thick accretion disk in a hot and luminous high-mass binary.\\

\section{Conclusion}

Millimeter and centimeter observations show that $\grs$ is at a kinematic distance from the Sun of $12.5 \plusoumoins 1.5 \kpc$, behind the core of a molecular cloud at $9.4 \plusoumoins 0.2 \kpc$. The column density of molecular gas $N(\mbox{H}_{2}) = 1.5 \plusoumoins 0.1 \times 10^{22} \cmdeux$, combined with the $\lambdavingtetun$ $\hun$ absorption, imply a total column density $\nh = 4.7 \plusoumoins 0.2 \times 10^{22} \cmdeux$, which corresponds to a visual absorption in the line of sight $\Av = 26.5 \plusoumoins 1 \mags$.\\

At a distance of $12.5 \kpc$ the hard X-ray transient $\grs$ often becomes the most powerful X-ray emitter in the Galaxy. The X-ray light curve observed since its discovery in 1992 (Paciesas et al. 1995) shows that, for recurrent periods that last several months it is one of the brighter sources of the sky at energies $\geq$ 20 keV. At $12.5 \kpc$ its X-ray luminosity is $\sim 3 \times 10^{38} \ergs$, which is indicative of super-Eddington accretion for a collapsed object of stellar mass.\\

The infrared counterpart of the source presents short- and long-term variations. We have observed a change of one magnitude over an interval of 24 hours, and changes of two magnitudes over intervals of months. No periodicity was detected. The similarity at infrared wavelengths between $\grs$ and $\ss$ suggests that both sources of relativistic jets are systems of similar nature. This similarity, together with the time variability of the infrared counterpart of $\grs$, indicates that a large fraction of the infrared emission may come from an accretion disk. Following the analogy with $\ss$, $\grs$ would be a collapsed object (neutron star or black hole) with a thick accretion disk in a high-mass and luminous binary system.

\begin{acknowledgements} We thank S. Massey for the infrared images obtained with the CFHT, T. Geballe and P. Charles (UKIRT Service program) for the K-magnitude of $\grs$ on 19 May 1995. We acknowledge helpful conversations with Christian Motch and Christian Gouiffes. We also thank John Simmons for reading the manuscript, and the anonymous referee for helpful suggestions. \end{acknowledgements}


\end{document}